# Tuning the metal-insulator transition in manganite films through surface exchange coupling with magnetic nanodots

T.Z. Ward[1,†], Z. Gai[1,2], X.Y. Xu[1], H.W.Guo[1,4], L.F. Yin[3], J. Shen[3,4,*]

[1]Materials Science and Technology Division, Oak Ridge National Laboratory, Oak Ridge, Tennessee 37830, USA
[2]Center for Nanophase Materials Sciences Division, Oak Ridge National Laboratory, Oak Ridge, Tennessee 37830, USA
[3]Dept. of Physics, Fudan University, Shanghai 200433, China
[4]Dept. of Physics and Astronomy, The University of Tennessee, Knoxville, Tennessee 37996, USA

[†] wardtz@ornl.gov
[*] shenj5494@fudan.edu.cn

**Abstract**
In strongly correlated electronic systems, such as manganites, the global transport behavior depends sensitively on the spin ordering, whose alteration often requires a large external magnetic field. Here we show that the spin ordering in manganites can be easily controlled by exchange field across the interface between a ferromagnet and a manganite. By depositing isolated ferromagnetic nanodots on a manganite thin film, we find that it is possible to overcome dimensionality and strain effects to raise the metal-insulator transition (MIT) temperature by over 200 K and increase the magnetoresistance by 5000%. The MIT temperature can also be tuned by controlling the density of the magnetic nanodots which indicates that the formation process of electronic phase separation can be controlled by the presence of magnetic nanodots.

Complex electronic systems exhibit a wide range of fascinating and potentially useful behaviors such as, colossal magnetoresistance (CMR), multiferroicity and high $T_c$ superconductivity. Thin films of these materials are particularly interesting for two reasons: 1) The physical parameters such as strain and dimensionality can be easily accessed to control the physical properties of the system; 2) Many practical applications require the materials to be prepared in thin film form. However, in many cases, the reduced dimensionality and strain effects can act against long range ordering and lead to critical temperatures that are too low for practical applications or lead to a loss of exotic properties altogether [1-8]. Restoring the long range order often requires impractically high external fields.

A typical example can be found in thin films of $La_{1-x}Ca_xMnO_3$ (LCMO) where the metal-insulator transition (MIT) temperature and CMR behavior can be reduced or destroyed depending on A-site doping and selected substrate [1,2,9,10]. The most common explanations given for this behavior is that the phase transition to ferromagnetic metal (FMM) is suppressed due to competing parasitic phases [9], bandwidth shrinkage arising from Mn-O-Mn bond deformations [3], and perturbation of spin alignment resulting from strain [11]. While the phase frustration cannot be overcome and the films remain as an insulator even under a high external field of 9 T, we show that depositing a layer of isolated magnetic Fe nanodots atop the film surface converts the system into metallic with a high MIT temperature. Experimental evidence clearly indicates that exchange coupling between the Fe nanodots' spins and the underlying Mn spins in the film plays the most critical role. Moreover, due to the highly intertwined nature of the energetics that drive electronic phase separation in manganites [12-17], the percolative MIT temperature is shown to be tunable by varying the density of the Fe nanodots.

For this study, we have grown epitaxial single crystal $La_{0.7}Ca_{0.3}MnO_3$ films of 20 nm thickness on $LaAlO_3$ (001) substrates having a miscut angle of < 0.1 degree and a nominal lattice mismatch to film of ~ -1.8%. The growth temperature was set at 1050 K with an $O_2$ growth pressure of $1\times10^{-3}$ Torr. Unit cell by unit cell growth was monitored by reflection high energy electron diffraction (RHEED) at a growth rate of ~1 unit cell per minute. X-ray diffraction (XRD) measurements were carried out on LCMO films and showed single phase epitaxial coverage with finite thickness peaks on the (002) line consistent with uniform 20nm thickness. Fe or Cu with nominal thickness of 1.0 nm were then thermally evaporated on the film surface at a substrate temperature ~ 320 K in ultrahigh vacuum condition. *Ex-situ* atomic force microscopy was used to confirm the formation of Fe or Cu nanodots on the LCMO surface. Those samples to be used for magnetic and transport characterizations were capped with an e-beam evaporated NaCl protecting layer before removing from vacuum. Magnetization vs temperature scans showed no evidence of either the Morin or Verwey transition on Fe nanodot treated films which confirms that iron oxidation was sufficiently arrested using this capping method. Transport measurements were performed on a Quantum Design Physical Property Measurement System (PPMS) using the 4-probe method with magnetic field directed perpendicular to the sample surface. An external Keithley 2400 with a maximum measurable resistance of 1GΩ was also used to confirm high resistance measurements. Magnetization data were taken using a Quantum Design Magnetic Properties Measurement system MPMS-XL (7T).

In our model system of a 20 nm LCMO film, the metal-insulator transition (MIT) is almost completely suppressed with only small magnetoresistance at low temperatures [1,3]. As shown in Figure 1a, the resistivity of the LCMO films increase monotonically with decreasing temperature, showing no sign of a metal-insulator transition down to the lowest measured temperature. Even with the application of high magnetic fields of up to 9 T, the insulating behavior of the film is preserved and only shows a modest reduction of resistivity below 100 K. Further, the sharp increase in resistivity at 150 K does not appear to be associated with the onset of a purely charge-ordered (CO) phase, as even under a high magnetic field (9 T) there is no indication of CO melting. As discussed by Bibes *et al.*, this seems to be evidence that non-FM regions effectively reduce the ferromagnetic coupling in the metallic regions due to localization of some $e_g$ electrons [3]. If this is the case, it can be expected that coupling a strong ferromagnet to the film could act to influence spin alignment in the film thereby overcoming the contributions coming from the non-FM regions and help to open the $e_g$ conduction channel.

The robust insulating behavior of the thin LCMO film, while showing little response to external magnetic field, changes dramatically when a layer of isolated magnetic iron nanodots is deposited on the surface. Figure 1b shows the temperature-dependent resistivity of the Fe nanodots capped LCMO film. The Fe nanodots have an average diameter of 29 nm and an average height of 12 nm, with an average distance to nearest neighbor of 80 nm. The typical morphology of the Fe dots can be seen in the inset of figure 1c. With the addition of the Fe nanodots, not only is the system converted into a metallic phase at low temperatures, but the metal-insulator transition temperature of ~255 K is similar to the highest recorded bulk values for this doping of LCMO. Further, the system exhibits a large magnetoresistive response upon application of a magnetic field.

The application of magnetic nanodots increased the MIT temperature by ~200 K and produced a 5000% increase in magnetoresistive response.

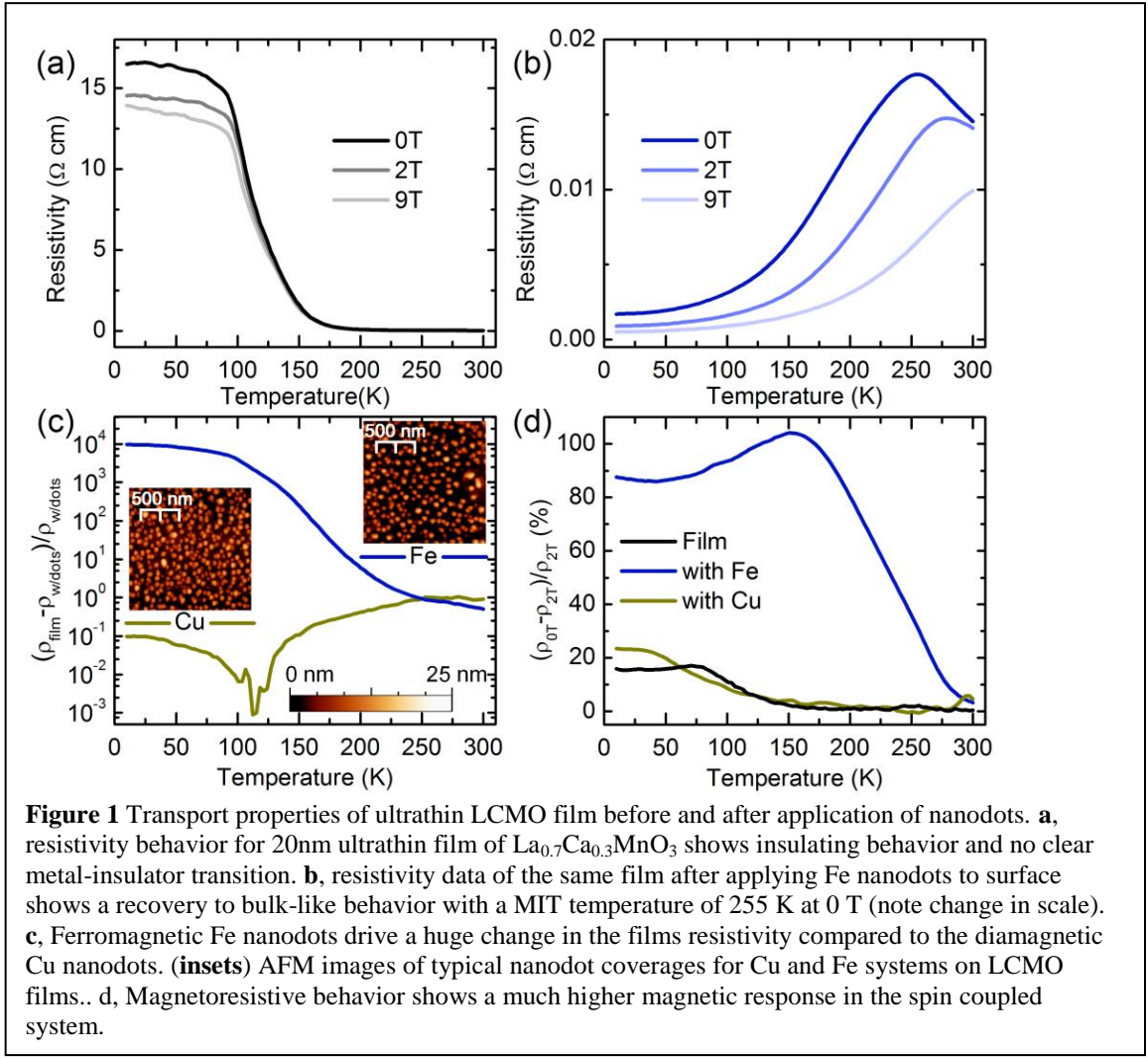

**Figure 1** Transport properties of ultrathin LCMO film before and after application of nanodots. **a**, resistivity behavior for 20nm ultrathin film of $La_{0.7}Ca_{0.3}MnO_3$ shows insulating behavior and no clear metal-insulator transition. **b**, resistivity data of the same film after applying Fe nanodots to surface shows a recovery to bulk-like behavior with a MIT temperature of 255 K at 0 T (note change in scale). **c**, Ferromagnetic Fe nanodots drive a huge change in the films resistivity compared to the diamagnetic Cu nanodots. (**insets**) AFM images of typical nanodot coverages for Cu and Fe systems on LCMO films.. d, Magnetoresistive behavior shows a much higher magnetic response in the spin coupled system.

It is well known that electronic phase separation commonly exists in manganite systems, and the corresponding MIT is considered percolative in nature [18-20]. With this in mind, one possible explanation for the observed transformation from an insulating state in the as-grown film to the metallic state in the nanodot treated film is that the addition of any metallic structures on the surface will simply act to connect the underlying metallic regions in the film thereby completing a percolation network. To test this, we observed the effects of non-magnetic Cu nanodots having average diameter of 26 nm, average height of 10 nm and an average nearest neighbor distance of 57 nm. As shown in figure 1c, the Cu nanodots have almost no influence on the overall resistive behavior of the LCMO film while the Fe nanodots produce a huge relative change in the film's resistivity, particularly at low temperatures. We also find that the magnetoresistance is greatly enhanced in the LCMO with Fe nanodots (figure 1d). Both the as-grown LCMO film and the Cu nanodots capped film exhibit similar, modest magnetoresistances of ~20% below 100 K. However, the Fe nanodots capped film shows a much stronger response—peaking at over 100% and covering a much broader temperature range. This vast improvement in the range of temperature response is also very interesting as most LCMO film and bulk systems tend to have a much narrower peak response window [21,22]. Further, even though the nanodot growth temperatures are too low for metal interdiffusion it is still important to note that the observed changes to resistive behavior do not appear to come from mixing of Cu or Fe into the LCMO film where they would replace Mn ions, as both of these substitutions have been shown to decrease the MIT temperature [23,24].

The fact that nonmagnetic Cu nanodots have almost no influence on the overall resistive behavior of the LCMO film is a strong indication that the magnetic interaction between the Fe nanodots and the LCMO film plays the critical role in converting the system into a metallic state. This magnetic interaction cannot be solely magnetostatic in nature, because the stray field from the Fe nanodots can at most reach ~ 2 T; and as observed in figure 1a, the uncapped LCMO film cannot be changed into a metallic state even under a 9 T external field. Therefore, we can speculate that the interaction between the Fe nanodots and the LCMO film is driven by the magnetic exchange interaction where the enormous exchange field drives alignment between Mn and Fe spins. Because all of the spins are ferromagnetically aligned within each Fe nanodot, the Mn spins in the LCMO regions underneath the Fe nanodots become ferromagnetically aligned as well. At the 0.3 Ca doping in bulk samples, such a ferromagnetic alignment of Mn spins will lead to a metallic state in the corresponding regions and eventually form conducting channels when the metallic regions become large and dense enough.

This scenario is supported by magnetization measurements of the system. Using a Quantum Design Magnetic Properties Measurement System (MPMS), we measured the magnetic response vs field of both as-grown and Fe nanodot capped LCMO films with magnetic field parallel and perpendicular to the surface at 150 K. In the as-grown ultrathin films, easy axis of magnetization is fairly balanced which is expected from previous studies [25,26] (figure 2a). The application of Cu nanodots had no noticeable effect and, similar to the film, showed no strong preference toward an easy axis of magnetization. However, after depositing Fe nanodots, the response to the parallel magnetic field increased substantially while the response to the perpendicular field decreased (figure 2b). When we compare the percent difference between the magnetization before and after Fe nanodot coverage, we see that the increase in the in-plane response is accompanied by a corresponding decrease in the out-of-plane response (figure 2c). Considering the fact that Fe nanodots should have an in-plane easy axis due to shape anisotropy, the drop in the out-of-plane response can be directly attributed to the modification of the LCMO film's magnetic state by spin coupling to the Fe nanodots, where the strong in-plane component from the Fe acts to align the spins in the film. This substantial drop in response also confirms that the magnetization measurements are not simple superpositions of Fe and LCMO signals. This was further confirmed by growing Fe nanodots on a bare LAO substrate using the same methods as discussed above; the saturation magnetization was over an order less than the observed increase in the in-plane measurements (figure 2 inset). The Fe nanodot system shows a moment of 1.9 $\mu_B$ at 10 K which is similar to a comparable volume of pure bcc Fe's moment of 2.15 $\mu_B$ and is considerably higher than reported values of 0.94 $\mu_B$ for surface oxidized Fe nanoparticles [27]. This suggests that while some very small amount of oxidation may be present in the nanoclusters on LCMO it does not substantially reduce the moment.

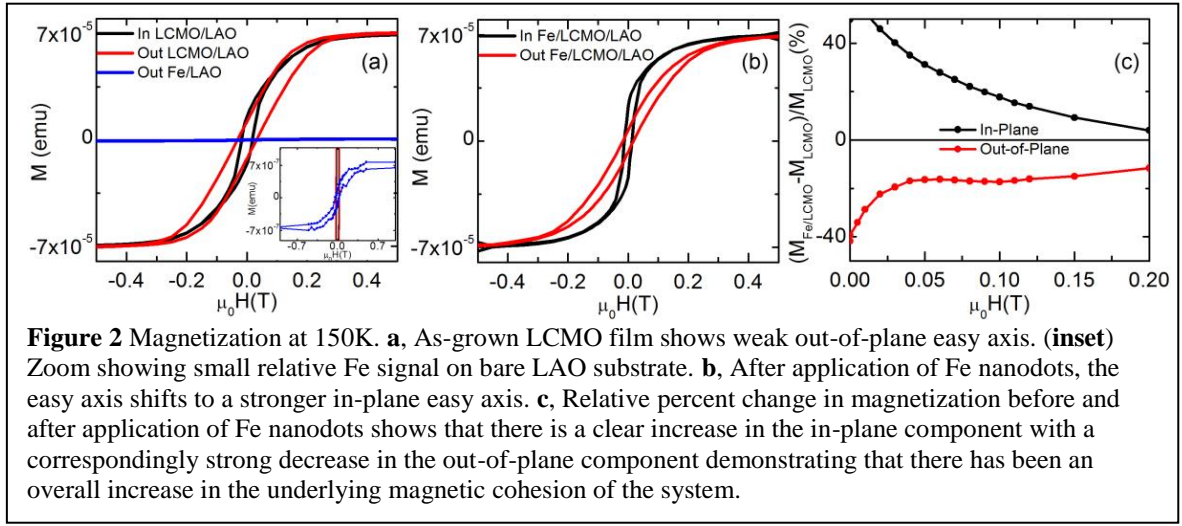

**Figure 2** Magnetization at 150K. **a**, As-grown LCMO film shows weak out-of-plane easy axis. (**inset**) Zoom showing small relative Fe signal on bare LAO substrate. **b**, After application of Fe nanodots, the easy axis shifts to a stronger in-plane easy axis. **c**, Relative percent change in magnetization before and after application of Fe nanodots shows that there is a clear increase in the in-plane component with a correspondingly strong decrease in the out-of-plane component demonstrating that there has been an overall increase in the underlying magnetic cohesion of the system.

Here we offer an explanation where the as-grown LCMO thin film is locked in an insulating phase due to its low dimensionality. This may either be caused by Mn-O-Mn bond distortion or some other mechanism that results in spin frustration which disallows a phase transition to a long range percolative FMM phase. Figure 3a shows a representation of this where the Mn spins are only weakly correlated and as such are only able to organize small regions of the FMM phase at low temperature. Upon application of Fe nanodots, the spins are locally ordered near the dot locations and are thus able to align themselves and undergo a phase transition to FMM which opens a percolative conduction channel through the film (figure 3b). In this scenario, the metal-insulator transition temperature is greatly increased with a corresponding improvement in the magnetoresistance. Further, the slight lowering of resistivity at temperatures above $T_C$ could be an indication that small polaron conduction is more easily mediated in the presence of the Fe nanodots; further work needs to be done to separate out the exact contributions from each conduction mechanism.

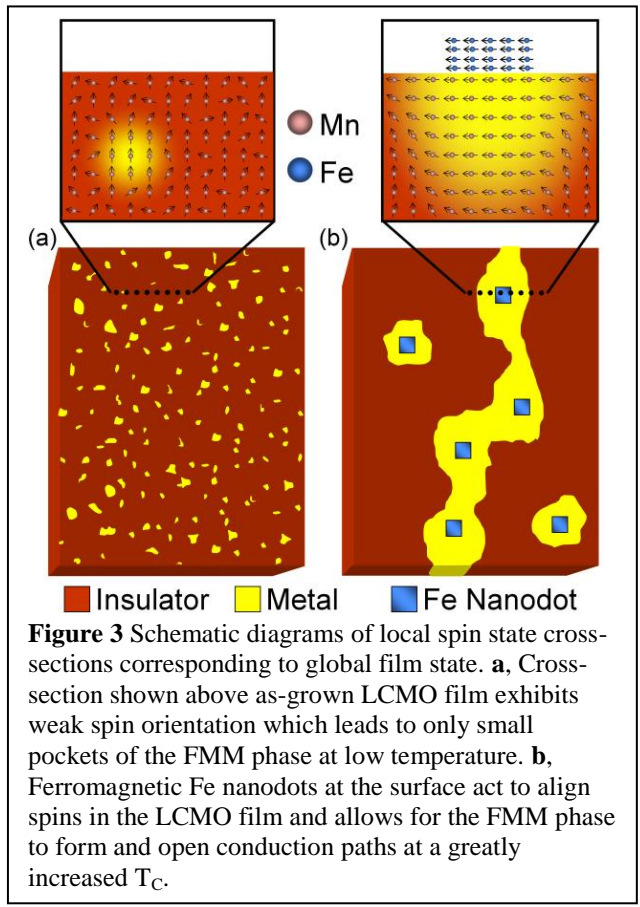

**Figure 3** Schematic diagrams of local spin state cross-sections corresponding to global film state. **a**, Cross-section shown above as-grown LCMO film exhibits weak spin orientation which leads to only small pockets of the FMM phase at low temperature. **b**, Ferromagnetic Fe nanodots at the surface act to align spins in the LCMO film and allows for the FMM phase to form and open conduction paths at a greatly increased $T_C$.

One of the most intriguing aspects of these findings is the possibility of tuning transport behaviors in these strongly correlated materials by simply varying the density or magnetic strength of the nanodots applied at the surface. As seen in figure 4, reducing the density of the Fe nanodots on the LCMO surface can have a dramatic effect on the transport properties. Here, the nanodots have an average diameter of 43 nm and an average height of 20 nm, with an average distance to nearest neighbor of 150 nm which is about 1/3 the density applied to the film in figure 1. At this lower density, the film recovers its magnetoresistive behavior and shows a strong response to applied magnetic fields, however the MIT temperature is decreased by about 135 K from the higher density system. This is consistent with the model discussed above where the insulating phase is pushed to a FMM transition. The lower transition temperature arises from the fact that the larger nearest neighbor distances between nanodots require a lower temperature for the percolation channels to open. While more detailed studies are certainly needed, this finding suggests that indeed resistive behaviors can be successfully controlled by tuning the parameters of the nanodot system driving the exchange coupling.

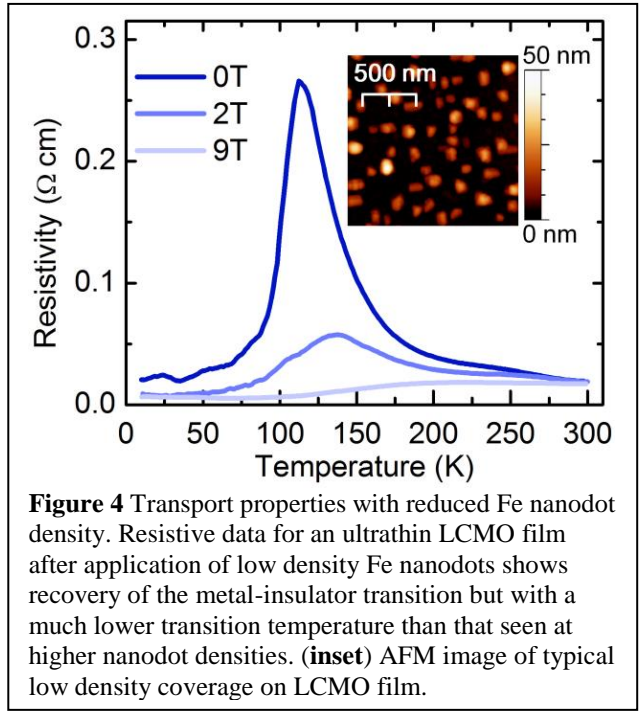

**Figure 4** Transport properties with reduced Fe nanodot density. Resistive data for an ultrathin LCMO film after application of low density Fe nanodots shows recovery of the metal-insulator transition but with a much lower transition temperature than that seen at higher nanodot densities. (**inset**) AFM image of typical low density coverage on LCMO film.

The transport and magnetization measurements show that it is possible to tune spin alignment in correlated electronic systems by spin coupling a ferromagnet at the surface. In the extreme example given in this work, even a system whose order parameters are completely unbalanced due to dimensionality and strain effects can be tuned back to its emergent state by adjusting the spin parameter through ferromagnetic nanodot densities. Importantly, this technique offers a new means to make specific investigations of spin contributions on emergent phenomena in complex materials. With the strong electronic correlation in complex materials, where spin, charge, orbital and lattice degrees of freedom overlap, this method should also find uses in materials ranging from colossal magnetoresistors to high $T_C$ superconductors in helping to tune spin dynamics. Finally, this method could act to push several complex material systems into working temperature ranges for new devices.

**Acknowledgments**: This effort was supported by the US DOE Office of Basic Energy Sciences, Materials Sciences and Engineering Division, through the Oak Ridge National Laboratory. We also acknowledge partial funding supports from the US DOE Office of Basic Energy Sciences, Scientific User Facilities Division, the US DOE grant DE-SC0002136, and the National Basic Research Program of China (973 Program) under the grant No. 2011CB921801.